\input harvmac
\input graphicx

\def\Title#1#2{\rightline{#1}\ifx\answ\bigans\nopagenumbers\pageno0\vskip1in
\else\pageno1\vskip.8in\fi \centerline{\titlefont #2}\vskip .5in}
%

%
%
\ifx\includegraphics\UnDeFiNeD\message{(NO graphicx.tex, FIGURES WILL BE IGNORED)}
\def\figin#1{\vskip2in}
\else\message{(FIGURES WILL BE INCLUDED)}\def\figin#1{#1}
\fi
\def\Fig#1{Fig.~\the\figno\xdef#1{Fig.~\the\figno}\global\advance\figno
 by1}
%
%
%
%
\def\Ifig#1#2#3#4{
\goodbreak\midinsert
\figin{\centerline{
\includegraphics[width=#4truein]{#3}}}
\narrower\narrower\noindent{\footnotefont
{\bf #1:}  #2\par}
\endinsert
}
%
%
\font\ticp=cmcsc10

\def\calo{{\cal O}}
\def\calh{{\cal H}}
\def\ehat{{\hat e}}

\def\limrho{\buildrel \rho\rightarrow \pi/2 \over \longrightarrow}

\def\roughly#1{\mathrel{\raise.3ex\hbox{$#1$\kern-.75em\lower1ex\hbox{$\sim$}}}}

\def\gbdel{{G_{B\partial}}}
\def\mthsu{\mathsurround=0pt  }
\def\leftrightarrowfill{$\mthsu \mathord\leftarrow\mkern-6mu\cleaders
  \hbox{$\mkern-2mu \mathord- \mkern-2mu$}\hfill
  \mkern-6mu\mathord\rightarrow$}
 \def\overleftrightarrow#1{\vbox{\ialign{##\crcr\leftrightarrowfill\crcr\noalign{\kern-1pt\nointerlineskip}$\hfil\displaystyle{#1}\hfil$\crcr}}}
\def\vecm{{\vec m}}
\def\vecx{{\vec x}}
\def\veck{{\vec k}}
\def\onl{{\omega_{nl}}}
\def\onz{{\omega_{n0}}}
\def\nlm{{nl\vecm}}
\overfullrule=0pt
%
%
\lref\Mald{
  J.~M.~Maldacena,
  ``The Large N limit of superconformal field theories and supergravity,''
  Adv.\ Theor.\ Math.\ Phys.\  {\bf 2}, 231 (1998)
  [Int.\ J.\ Theor.\ Phys.\  {\bf 38}, 1113 (1999)]
  [arXiv:hep-th/9711200].
}
\lref\PolKITP{
  J.~Polchinski,
  ``Emergent Spacetime from AdS/CFT,''
  talk, The Harmony of Scattering Amplitudes, KITP (2011).
  {\it http://online.itp.ucsb.edu/online/qcdscat11/polchinski/} 
}
\lref\Nonlocvscomp{
  S.~B.~Giddings,
  ``Nonlocality versus complementarity: A Conservative approach to the information problem,''
  Class.\ Quant.\ Grav.\  {\bf 28}, 025002 (2011)
  [arXiv:0911.3395 [hep-th]].
}
\lref\BuLu{
  C.~P.~Burgess and C.~A.~Lutken,
  ``Propagators and Effective Potentials in Anti-de Sitter Space,''
  Phys.\ Lett.\  B {\bf 153}, 137 (1985).
}
\lref\AnMa{
  T.~Andrade, D.~Marolf,
 ``AdS/CFT beyond the unitarity bound,''
[arXiv:1105.6337 [hep-th]].
}
\lref\DHFr{
  E.~D'Hoker, D.~Z.~Freedman, S.~D.~Mathur, A.~Matusis and L.~Rastelli,
  ``Graviton and gauge boson propagators in AdS(d+1),''
  Nucl.\ Phys.\  B {\bf 562}, 330 (1999)
  [arXiv:hep-th/9902042].
}
\lref\FeffGr{
  C.~Fefferman and R.~Graham,
  ``Conformal invariants,''
  in {\it \'Elie Cartan et les Math\'emati\-ques d'Aujourdui}, Ast\'erisque, 95 (1985)
}
\lref\Erice{
  S.~B.~Giddings,
  ``The gravitational S-matrix: Erice lectures,'' Subnucl.\ Ser.\  {\bf 48}, 93 (2013), 
[arXiv:1105.2036 [hep-th]].
}\lref\EaGi{
  D.~M.~Eardley and S.~B.~Giddings,
  ``Classical black hole production in high-energy collisions,''
  Phys.\ Rev.\  D {\bf 66}, 044011 (2002)
  [arXiv:gr-qc/0201034].
}
\lref\forty{
  S.~B.~Giddings,
  ``Is string theory a theory of quantum gravity?,''
[arXiv:1105.6359 [hep-th]], Found.\ Phys.\  {\bf 43}, 115 (2013).
}
\lref\Coleman{
  S.~Coleman,
  ``Acausality,''
  In {\sl Erice 1969, Ettore Majorana School On Subnuclear Phenomena,} New York 1970, 282-327.}
\lref\FiKa{
  A.~L.~Fitzpatrick, J.~Kaplan,
  ``Scattering States in AdS/CFT,''
[arXiv:1104.2597 [hep-th]].
}
\lref\GGP{
 M.~Gary, S.~B.~Giddings, J.~Penedones,
  ``Local bulk S-matrix elements and CFT singularities,''
Phys.\ Rev.\  {\bf D80}, 085005 (2009).
[arXiv:0903.4437 [hep-th]].
}
\lref\GaGi{
  M.~Gary and S.~B.~Giddings,
  ``The flat space S-matrix from the AdS/CFT correspondence?,''
  Phys.\ Rev.\  D {\bf 80}, 046008 (2009)
  [arXiv:0904.3544 [hep-th]].
}
\lref\GiSr{
  S.~B.~Giddings and M.~Srednicki,
  ``High-energy gravitational scattering and black hole resonances,''
  Phys.\ Rev.\  D {\bf 77}, 085025 (2008)
  [arXiv:0711.5012 [hep-th]].
}
\lref\GiPo{
 S.~B.~Giddings and R.~A.~Porto,
  ``The gravitational S-matrix,''
  Phys.\ Rev.\  D {\bf 81}, 025002 (2010)
  [arXiv:0908.0004 [hep-th]].
}
\lref\PolS{
  J.~Polchinski,
  ``S-matrices from AdS spacetime,''
  arXiv:hep-th/9901076.
}
\lref\SussS{
  L.~Susskind,
 ``Holography in the flat space limit,''
  arXiv:hep-th/9901079.
}
\lref\GKP{
  S.~S.~Gubser, I.~R.~Klebanov and A.~M.~Polyakov,
  ``Gauge theory correlators from non-critical string theory,''
  Phys.\ Lett.\  B {\bf 428}, 105 (1998)
  [arXiv:hep-th/9802109].
}
\lref\WittAdS{
  E.~Witten,
  ``Anti-de Sitter space and holography,''
  Adv.\ Theor.\ Math.\ Phys.\  {\bf 2}, 253 (1998)
  [arXiv:hep-th/9802150].
}
\lref\Katzetal{
  A.~L.~Fitzpatrick, E.~Katz, D.~Poland and D.~Simmons-Duffin,
  ``Effective Conformal Theory and the Flat-Space Limit of AdS,''
  arXiv:1007.2412 [hep-th].
}
\lref\FSS{
  S.~B.~Giddings,
  ``Flat-space scattering and bulk locality in the AdS/CFT  correspondence,''
  Phys.\ Rev.\  D {\bf 61}, 106008 (2000)
  [arXiv:hep-th/9907129].
}
\lref\BSM{
  S.~B.~Giddings,
  ``The Boundary S matrix and the AdS to CFT dictionary,''
Phys.\ Rev.\ Lett.\  {\bf 83}, 2707-2710 (1999).
[hep-th/9903048].
}
\lref\HPPS{
  I.~Heemskerk, J.~Penedones, J.~Polchinski and J.~Sully,
  ``Holography from Conformal Field Theory,''
  JHEP {\bf 0910}, 079 (2009)
  [arXiv:0907.0151 [hep-th]].
}
\lref\ReSi{
  M.~Reed and B.~Simon,
  ``Methods Of Mathematical Physics. Vol. 3: Scattering Theory,''
  {\it  New York, USA: Academic (1979) 463p}.
}
\lref\BKL{
  V.~Balasubramanian, P.~Kraus and A.~E.~Lawrence,
  ``Bulk versus boundary dynamics in anti-de Sitter space-time,''
  Phys.\ Rev.\  D {\bf 59}, 046003 (1999)
  [arXiv:hep-th/9805171].
}
\lref\Raju{
  S.~Raju,
  ``Generalized Recursion Relations for Correlators in the Gauge-Gravity Correspondence,''
  Phys.\ Rev.\ Lett.\  {\bf 106}, 091601 (2011).
}
\lref\FiKaUnitarity{
  A.~L.~Fitzpatrick and J.~Kaplan,
  ``Unitarity and the Holographic S-Matrix,''
[arXiv:1112.4845 [hep-th]].
}
\lref\FitzpatrickHU{
  A.~L.~Fitzpatrick and J.~Kaplan,
  ``Analyticity and the Holographic S-Matrix,''
[arXiv:1111.6972 [hep-th]].
}
\lref\DiasSS{
  O.~J.~C.~Dias, G.~T.~Horowitz and J.~E.~Santos,
  ``Gravitational Turbulent Instability of Anti-de Sitter Space,''
[arXiv:1109.1825 [hep-th]].
}

\lref\FitzpatrickIA{
  A.~L.~Fitzpatrick, J.~Kaplan, J.~Penedones, S.~Raju and B.~C.~van Rees,
  ``A Natural Language for AdS/CFT Correlators,''
JHEP {\bf 1111}, 095 (2011).
[arXiv:1107.1499 [hep-th]].
}

\lref\PenedonesS{J. Penedones, talk at Strings 2015.}

\lref\Polcomm{J. Polchinski, comment on talk by E.~Silverstein at the workshop Black Holes, Complementarity, Fuzz, or Fire, KITP, UCSB, August 2013,\hfil \break http://online.kitp.ucsb.edu/online/fuzzorfire\_m13/silverstein/\ .}

\lref\ChristodoulouKlainerman{D.~Christodoulou and S.~Klainerman,
 ``The Global nonlinear stability of the Minkowski space,''
Princeton University Press, Princeton, 1993.
}

\lref\BizonRostworowski{
  P.~Bizon and A.~Rostworowski,
  ``On weakly turbulent instability of anti-de Sitter space,''
Phys.\ Rev.\ Lett.\  {\bf 107}, 031102 (2011).
[arXiv:1104.3702 [gr-qc]].
}

\lref\AndersonAX{
  M.~T.~Anderson,
 ``On the uniqueness and global dynamics of AdS spacetimes,''
Class.\ Quant.\ Grav.\  {\bf 23}, 6935 (2006).
[hep-th/0605293].
}

\Title{
\vbox{\baselineskip12pt
}}
{\vbox{\centerline{Constraints on a fine-grained AdS/CFT correspondence}
}}
\centerline{{\ticp Mirah Gary\footnote{$^\dagger$}{Email address: mgary@hep.itp.tuwien.ac.at } and
Steven B. Giddings\footnote{$^\ast$}{Email address: giddings@physics.ucsb.edu}  
} }
\centerline{\sl Department of Physics}
\centerline{\sl University of California}
\centerline{\sl Santa Barbara, CA 93106}
\vskip.08in
\centerline{\bf Abstract}
For a boundary CFT to give a good approximation to the bulk flat-space S-matrix, a number of conditions need to be satisfied: some of those are investigated here.  In particular, one would like to identify an appropriate set of approximate asymptotic scattering states, constructed purely via boundary data.  We overview, elaborate, and simplify obstacles encountered with existing proposals for these.  Those corresponding to normalizable wavefunctions undergo multiple interactions; we contrast this situation with that needed for a flat-space LSZ treatment.  Non-normalizable wavefunctions can have spurious interactions, due either to power-law tails of wavepackets or  to their non-normalizable behavior, which obscure S-matrix amplitudes we wish to extract; although in the latter case we show that 
such gravitational interactions can be finite, as a result of gravitational red shift.
We outline an illustrative construction of arbitrary normalizable wavepackets from boundary data, that also yields such spurious interactions.  Another set of non-trivial questions regard the form of unitarity relations for the bulk S-matrix, and in particular its normalization and multi-particle cuts.  These combined constraints, together with those found earlier on boundary singularity structure needed for bulk momentum conservation and other physical/analytic properties, are a non-trivial collection of obstacles to surmount if a fine-grained S-matrix, as opposed to a coarse-grained construction, is to be defined purely from boundary data.

\vfil
\Date{}

\newsec{Introduction}

A complete quantum theory of gravity must address non-perturbative questions.  Particularly puzzling among these regard the dynamics of ultraplanckian scattering, and of quantum cosmology.  While perturbative string theory has addressed other problems of quantum gravity, particularly perturbative infinities and perturbative nonrenormalizability, and while significant non-perturbative structures have been investigated in the theory, it remains to be seen what exactly it says about these non-perturbative phenomena.

If string theory is to describe nature, a complete non-perturbative formulation of it is therefore certainly needed.  This is a long-standing problem, which many have advocated is solved by the AdS/CFT correspondence\refs{\Mald}.\foot{Quoting Polchinski\refs{\PolKITP}, description of strongly coupled gauge theory phenomena is only AdS/CFT's ``hobby, ... its  real job is to provide a non-perturbative construction of quantum gravity.''}

However, despite much discussion, the non-trivial question of how a lower-dimensional boundary theory can capture the full higher-dimensional bulk physics has remained mysterious.  As we will review, perhaps the best prospect is to find a close approximation to the bulk S-matrix from CFT data (for relevant discussion, see \refs{\PolS\SussS\FSS\GGP\GaGi\HPPS\Katzetal-\FiKa}), although a more complete description of quantum gravity would seem to also require some approximate notion of local bulk observables\refs{\forty}.

As we will describe further, a number of non-trivial conditions must be satisfied in order to extract an approximate bulk S-matrix with the correct physical properties; it remains to be seen whether this can be done.  A possible alternative\refs{\FSS,\GaGi,\forty} is that AdS/CFT provides a kind of coarse-grained correspondence, in which the boundary theory may be somewhat analogous to an effective theory, without capturing all the fine-grained detail of the bulk theory.  Indeed, while a number of interesting descriptions of boundary phenomena have arisen in exploring the correspondence, these do not seem to depend on a fine-grained match.
And, if there were such a detailed correspondence, one would like to understand how the various necessary conditions are satisfied by the boundary theory -- which would explain the proposed ``miracle'' of holography.

Particularly mysterious is the question of higher-dimensional locality.  While we do not expect a bulk theory with exact locality -- indeed, some nonlocality is apparently needed to resolve the unitarity crisis of ultraplanckian scattering\refs{\Nonlocvscomp} -- it should behave approximately locally in familiar low-energy contexts, in order to describe the physical phenomena we observe.  In field theory, locality can be formulated in terms of microcausality -- commutativity of local observables outside the light cone.  But, without such a construction of local observables, one can also probe locality through properties of the S-matrix.  These include clustering and macrocausality\refs{\Coleman}, which state that at large separations amplitudes for processes decouple, and the leading order correction to this is single-particle exchange.  Another related property is the analytic structure of the S-matrix, and particularly its polynomiality.\foot{For further discussion of these questions see \refs{\GiPo,\Erice}.}  

Thus, we should seek a construction of an approximate bulk S-matrix with these and other necessary physical features, purely from boundary data.  The first question that arises is how such a candidate S-matrix emerges.  As we will outline in the next section, we need a description both of asymptotic scattering states, and of amplitudes of the right form connecting them.  There are non-trivial questions in formulating such scattering states from boundary data.  If they correspond to normalizable bulk states, as described in section three, the failure of particles to asymptotically separate interferes with attempts to isolate the amplitude for a single scattering in a given ``interaction region;'' instead, particles can undergo infinitely many interactions\GaGi.  

Alternatively, boundary sources can be turned on at finite times.  During such times, the bulk wavefunction is not normalizable, and can be thought of as describing infinitely many particles ``near infinity.''  This raises questions of how to formulate unitarity, and of other such spurious interactions between particles contaminating the amplitudes we want.  One approach to this, described in \refs{\GGP,\GaGi} and refined in section four, is to consider sources with compact support.  These, however, generically have power law tails, resulting from singular behavior at small momenta, and limiting their resolving power.

We are therefore led to confront the issues of non-compact boundary sources in section five.  We find that the gravitational interaction is special, in that it exhibits finite interactions between non-normalizable wavefunctions, due to gravitational redshift as the boundary of AdS is approached.  (In order to show this, we find and correct an error in the derivation of the graviton propagator given in \DHFr.)
Nonetheless, there remains the possibility of finite interactions confounding attempts to extract S-matrix amplitudes.  We construct a new type of state, which we call ``resonant'' wavepackets, illustrating this behavior, in section six.  Specifically, this construction allows us to create an {\it arbitrary} bulk wavepacket from a boundary source at finite time.  However, two such wavepackets that are designed to scatter in a specified interaction region will also generically have confounding interactions in an image interaction region, while the wavepackets are being built up.  While we have not quite proven a no-go theorem for construction of appropriate scattering states without spurious interactions, this discussion gives additional illustration of the general problem of isolating desired scattering amplitudes from those of such spurious scatterings, due to limitations on our control of bulk states via the boundary theory.

Another  critical property for a candidate bulk S-matrix is unitarity.  While it is commonly stated that boundary unitarity implies bulk unitarity, the origin of the unitarity condition for the bulk S-matrix, $S^\dagger S=1$, is not obvious from the boundary perspective.  We examine this question, assuming (in contrast to the preceding story) that there  is {\it some} construction of bulk scattering states in terms of boundary operators.  While factorization on a single-particle intermediate state has a parallel in the operator product expansion, perturbative unitarity also implies various non-trivial cuts, beginning with two-particle intermediate states.  We outline another mystery of a proposed fine-grained correspondence, which is the possible origin of such cuts, from the boundary perspective.  

This paper closes with a summary of such constraints on attempts to derive a bulk S-matrix from the boundary theory; these also include the ones described in \GGP, where it was shown that special kinds of singularities are needed in the boundary correlators in order to reproduce bulk momentum conservation and long-distance gravitational Born-scattering behavior.  The combined weight of these constraints raises serious questions about viability of the proposal to extract detailed bulk physics from the boundary theory.

\newsec{The question of a fine-grained correspondence and the flat S-matrix}

The most straightforward and common interpretation of the AdS/CFT correspondence is as an equivalence between quantum-mechanical theories describing bulk and boundary physics. Thus, the correspondence should be described as a map between the Hilbert spaces of the two interacting theories, 
\eqn\hilbmap{ M: \calh_B\rightarrow \calh_\partial}
that is one to one and onto,  and unitary. 

The Hilbert space $\calh_B$ is proposed to be that of interacting string theory with asymptotically AdS boundary conditions.  If this is the case, we expect certain kinds of states to be present in this Hilbert space.  In particular, consider the situation where the AdS radius is $R\sim 10^{10}$ light-year, and the string coupling $g$ is small but finite.  A complete bulk Hilbert space is then expected to include states that at a given time look like a pair of wavepackets for two particles, each with energy $3.5\ TeV$, separated by km-scale distances, and individually well-localized to scales much smaller than this separation.  Indeed, such a state is what is needed for a description of an LHC collision,\foot{Of course, here one also has spacetime dimension $D=4$, but if AdS/CFT is to be useful for defining string theory, we need a close parallel to the case of $AdS_5\times S^5$.} and we expect description of such a state to be essentially independent of a curvature radius of cosmological scales -- very sensitive measurements are needed to determine this vacuum curvature.  By giving the quantum amplitudes relating ingoing and outgoing states of this kind, AdS/CFT would then determine a very close approximation to the flat space S-matrix, or an inclusive version accounting for the usual soft photon/graviton divergences in $D=4$.  We will refer to the region in which we wish to isolate such scattering amplitudes, which is small as compared to the AdS radius, as the {\it interaction region.}

An essential question is whether this is indeed the case -- that there is a fine-grained correspondence \hilbmap\ permitting us to study localized bulk states and in particular the S-matrix in the flat-space limit.  Since one would like this to serve as a {\it definition} of string theory in AdS, in terms of the more-familiar boundary gauge theory, such a question needs to be addressed under the assumption that one ultimately only has direct definition and control of quantities in the boundary theory.   If the answer to this question is yes, then one would like to infer the properties of the boundary theory responsible for various behavior of the bulk theory; a particularly puzzling question is how familiar locality of the higher-dimensional bulk is (approximately) encoded in the boundary theory.

In a familiar QFT description via LSZ, elements of the flat-space S-matrix take the general form
\eqn\bulkS{S[\psi_i] = \int \prod_i\left[ dx_i \psi_i(x_i)\right] G_T(x_i)\ ,}
where the $\psi_i$ are scattering states living in the appropriate free Hilbert space for asymptotic particles, and $G_T(x_i)$ is an amputated Green function.  So, to give a good approximate match to such a description, one needs both a way of constructing, via the boundary theory, an appropriate space of scattering states, {\it and} a resulting $S[\psi_i]$ with certain properties needed to closely approximate familiar physics.  These include momentum conservation (perhaps up to accuracy $\sim 1/R$), and other observed features, such as long-distance Coulomb scattering behavior, {\it etc.}  Some of these features of the gravitational S-matrix are summarized in \refs{\Erice}.  Or, one might imagine that as an alternative the boundary theory directly defines quantities that closely approximate the bulk S-matrix, without the intermediary of describing precise scattering states.  

Part of the AdS/CFT conjecture, as elaborated by \refs{\GKP,\WittAdS,\PolS,\SussS}, is that quantities like \bulkS\ arise from correlators of boundary operators $\calo_i$ smeared against boundary sources $f_i$:
\eqn\bcorr{\left\langle \int \prod_i\left[db_i f_i(b_i)\calo_i(b_i)\right] \right\rangle_{CFT}\ ,}
where the $b_i$ denote boundary points. But, in the details of such a construction one encounters multiple questions.\foot{Note that while \BSM\ argues that this construction defines a ``boundary S-matrix" for AdS, at present we are inquiring whether the boundary theory approximates detailed data of a flat space S-matrix.}

In particular, detailed aspects of the relation of \bcorr\ to \bulkS\ can be investigated in the case where one assumes that one has a bulk field theory ({\it e.g.} supergravity) approximating string theory.  This theory can be used to construct boundary correlators, via the map proposed by Gubser, Klebanov, Polyakov, and Witten (GKPW)\refs{\GKP,\WittAdS}, and then one can ask whether bulk quantities such as \bulkS\ can be recovered purely from these boundary quantities, without additional assumptions about details of bulk physics.

Let us recall basic aspects of this construction.\foot{Many useful formulas can be found in \FSS, though we will convert to the conventions of \refs{\GaGi}.}  If $\phi$ is a bulk field of mass $m$, its bulk correlators determine correlators of the corresponding boundary operator $\calo$ through the formula
\eqn\boundcorr{\langle\calo(b)\cdots\rangle =  (2\Delta-d) R^{(d-1)/2}\lim_{\rho\rightarrow\pi/2} (\cos\rho)^{-\Delta}\langle \phi(\tau,\rho,\ehat)\cdots\rangle\ .}
Here, $b=(\tau,\ehat)$ denotes the boundary coordinates, and the full bulk metric takes the form
\eqn\glocoora{ds^2 = {R^2\over \cos^2\rho} ( -d\tau^2 + d\rho^2 +
\sin^2\rho \, d\Omega^2_{d-1} )\ .}
The conformal dimension $\Delta$ of the boundary operator is given by
\eqn\confdim{\Delta =  {d \pm\sqrt{d^2 + 4m^2R^2}\over 2}\ .}

Conversely, integrating $\calo(b)$ against a boundary source $f(b)$ produces a specific bulk wavefunction.  This can be determined from the bulk boundary propagator, which is a limit of the bulk Feynman propagator\foot{The Feynman prescription naturally arises when describing the bulk interactions encoded in \bulkS\ perturbatively.}  $G_B$,
\eqn\gbbdef{\gbdel(b',x) = (2\Delta-d) R^{(d-1)/2} \lim_{\rho'\rightarrow\pi/2}(\cos\rho')^{-\Delta}
G_B(x',x)\ ,}
as in \boundcorr.  Specifically, one finds
\eqn\bdywavepack{\psi_f(x) =\int db' f(b') G_{B\partial}(b',x)\ . }
For a general source $f$ this wavefunction is non-normalizable, with asymptotic behavior
\eqn\psinn{\psi_f(x)\limrho {(\cos\rho)^{d-\Delta}\over R^{(d-1)/2}} f(b)\ .}

Here we see the origin of part of the problem of extracting the S-matrix.  In QFT, one can give a particle-number current, which for a scalar is the Klein-Gordon current,
\eqn\KGcurr{j_\mu = i \phi^* \overleftrightarrow{\partial_\mu} \phi\ .}
For the wavefunction $\psi_f$, the particle number integrated over a surface of constant $\tau=\tau_0$, with surface element $d\Sigma^\mu = d^d x \sqrt{g_d} n^\mu$, 
\eqn\partnum{N=\int_{\tau=\tau_0} d\Sigma^\mu j_\mu\ ,}
is infinite if $f(\tau_0,\ehat)\neq0$ in some neighborhood on the boundary.  An explanation of this is that there are infinitely many $R$-sized volumes ``at infinity'' in AdS, and for a solution with behavior \psinn, these will contain infinite particle number.  If, in order to extract the S-matrix, one is trying to focus on a single scattering event of a pair of particles, in a specific $R$-sized interaction region which by convention we can take to be the center of AdS in coordinates \glocoora, there is a potential for contamination of the scattering amplitude from scatters of the infinitely many other particles.  Indeed, one might be concerned that generically one finds {\it infinite} scattering amplitudes\FSS, and that the sources $f_i(b_i)$ for scattering states consequently must be chosen with some care.

In the free theory, $g=0$, there is a natural way to circumvent this problem: take the source $f$ to only have support in the far past.  In that case, it produces a normalizable solution.  A basis for such solutions can be written\refs{\BuLu,\BKL}
\eqn\normmodes{\phi_{nl\vecm}(\vecx,\tau) = \chi_{nl}(\rho) Y_{l\vecm}(\ehat)
{e^{-i\onl\tau}\over \sqrt{2\onl}}= \phi_{nl\vecm}(\vecx){e^{-i\onl\tau}\over \sqrt{2\onl}} \ .}
Here $n$ is a principle quantum number and $\chi$ a radial wavefunction, $Y_{l\vecm}(\ehat)$ are spherical harmonics for the appropriate dimension, and further details are given  
in \FSS.  These have normal-mode frequencies
\eqn\normfreq{\omega_{nl} = \Delta + 2n+l \ .}

Sources can be constructed,\foot{See sec.~six for an explicit example.} with support purely in the far past, that create a general normalizable wavepacket
\eqn\gennorm{\psi(x) = \sum_{nl\vecm} c_{nl\vecm} \phi_{nl\vecm}(\vecx,\tau)\ .}
In the limit $R\rightarrow\infty$, with fixed bulk energy $\omega = \omega_{nl}/R$, the radial wavefunctions are well approximated by Bessel functions (see \FSS), and one can thus closely approximate an arbitrary and perfectly well-behaved flat-space wavepacket.   Notice an important property of such wavepackets: they are precisely periodic (up to an overall phase) in $\tau$ with period $2\pi$.  Thus, such a focussed wavepacket or its mirror image precisely refocusses at $\tau$ intervals separated by $\pm \pi$.  In terms of proper time at the center of AdS, the periodicity is $2\pi R$.

For small coupling, $g\ll 1$, one na\"\i vely expects merely a small, perturbative, departure from this story, thus allowing us to formulate the S-matrix for non-zero coupling.  Unfortunately, this expectation appears unfounded.

\newsec{Normalizable states and multiple interactions}

 \Ifig{\Fig\adscoll}{AdS behaves like a gravitational ``box'' of radius $R$; normalizable states correspond to particles confined to the box, which undergo multiple collisions.}{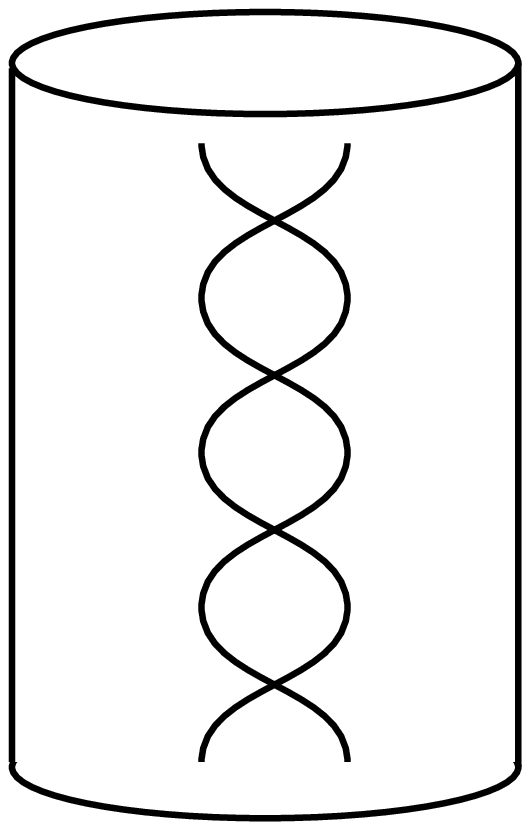}{2}

The basic issue with normalizable wavepackets is that they don't asymptotically separate; instead, due to the periodicity we have noted, two free particles that pass close to one another once will pass at the same distance an infinite number of times, as pictured in \adscoll.  This suggests that even a very small interaction does not have a small effect on the amplitude to go from a given initial state to a given final state.  Put differently, if one is trying to set up a physical situation with an incoming state like that for LHC collisions described above, and find the amplitude to scatter to a similar outgoing state, one needs to know that one can isolate the scattering amplitude for {\it this} collision, without important contributions from other past or future scatterings contaminating the amplitude.  

In an interacting field theory in flat space, a careful treatment of scattering theory, such as that of Haag-Ruelle or LSZ, is needed; let us explore the analog of the latter in AdS.  With nonzero coupling,
 \hilbmap\ would be a map between the Hilbert spaces of the interacting theories.   One would like to construct asymptotically two-particle scattering states of the kind described above.  If there are one particle states of the interacting theory, which we denote $|\Delta,n,l,\vecm\rangle$ with quantum numbers corresponding to those in \normmodes, a well-localized wavepacket corresponding to one of the particles can be chosen by taking an appropriate superposition,
 \eqn\onepart{|f\rangle = \sum_{n, l,\vecm} c_\nlm |\Delta, n, l,\vecm\rangle\ }
 The question is how to construct the appropriate two-particle normalizable state.  
 
 First, let us return to the {\it free} bulk theory, with fields normalized such that
 \eqn\fieldnorm{\langle\Delta, n, l,\vecm|\phi(x) |0\rangle = N(\Delta) \phi^*_\nlm(x)\ .}
 The state \onepart\ has wavefunction given by
 \eqn\onewave{\langle0| \phi(x) |f\rangle = N(\Delta) \psi_f(x)\ ,}
which is a superposition of basis solutions:
 \eqn\fwave{\psi_f(x)=\sum_\nlm c_\nlm \phi_\nlm(x)\ .}
 The state \onepart\ can be created using operators
 \eqn\create{\alpha_f(\tau) = {1\over N(\Delta)} \int_\tau d\Sigma^\mu \psi^*_f i {\overleftrightarrow{\partial_\mu}} \phi\ .}
 Then, the desired two-particle state of the free theory is
 \eqn\twopart{ \alpha^\dagger_{f_1} \alpha^\dagger_{f_2} |0\rangle\ .}

 In the interacting theory, with normalization \fieldnorm\ for the {\it interacting} (Heisenberg picture) field, in parallel with the LSZ procedure one may again use \create, with $\Delta$ corresponding to the single-particle mass in the interacting theory, to create a single-particle wavepacket.  In doing this, one takes the limit $\tau\rightarrow-\infty$ in $\alpha_f(\tau)$.  The reason for this is that the interacting field $\phi(x)$ can also create multi-particle states.  In conventional LSZ, the limit projects onto the one-particle component, and we will assume there is no problem with this here.  In the AdS context, this is also important because we want to consider normalizable states, without turning on sources at finite times.
 Thus, following the usual LSZ procedure, we might propose that \twopart\ is the desired two-particle state, where
 \eqn\alphadef{\alpha_f = \lim_{\tau\rightarrow -\infty} \alpha_f(\tau)\ .}

 We can diagnose the behavior of the interacting state \twopart\ through
 \eqn\shape{\langle0|\phi(x_1)\phi(x_2) \alpha^\dagger_{f_1} \alpha^\dagger_{f_2} |0\rangle\ ;}
 which should then give the desired localized wavepackets.  To check this via perturbation theory (connecting with the Feynman rules) the standard procedure is to switch to interaction picture.  The interaction-picture evolution operator is
\eqn\intevol{U_I(\tau_2-\tau_1) = T \exp\left\{-i\int_{\tau_1}^{\tau_2} dt H_I(\tau)\right\}\ ,}
where $H_I(\tau)$ is the interaction hamiltonian due to the coupling.  The interaction-picture field is
\eqn\ipfield{\phi_I(x) = U_I(\tau) \phi(x) U_I^\dagger(\tau)\ .}
Eq.~\shape\ then becomes (assuming equal times $=\tau$ for $x_1$, $x_2$)
\eqn\ipcorr{\langle0|\phi(x_1)\phi(x_2) \alpha^\dagger_{f_1} \alpha^\dagger_{f_2} |0\rangle= \lim_{\tau'\rightarrow-\infty}\langle0| U_I^\dagger(\tau) \phi_I(x_1)\phi_I(x_2) U_I(\tau-\tau') \alpha^\dagger_{I,f_1}(\tau') \alpha^\dagger_{I,f_2}(\tau') U_I(\tau')|0\rangle\ .}
Recall that the interaction-picture fields satisfy the free-field equations.  In flat space, appropriately chosen wavepackets will  experience interactions described by $U_I(\tau-\tau')$ only in our given interaction region; otherwise the wavepackets are widely separated and effectively non-interacting.  But, in AdS 
eq.~\ipcorr\ and its corresponding Feynman diagrams exhibit the problem: the particles incident from the infinite past experience the infinite number of interactions described by $U_I(\tau-\tau')$ and pictured in \adscoll.  Thus, if the $f_i$'s were chosen to produce the desired wavepackets in the free theory, they do not in general do so taking into account these interactions, which can furthermore produce more particles.
 
For this discussion we have assumed we can make use of field theory in the bulk.  If we are trying to formulate purely in the boundary theory amplitudes corresponding to transitions between states like \twopart, we might imagine that they are of the form \bcorr, with appropriate sources turned on in the far past and future.  But, there is no apparent reason we would have better control over the state appearing in the required form, with incoming localized wavepackets at a given finite time, if we indeed work directly with the boundary description of the state.

One recent suggestion\Katzetal\ is that the bulk S-matrix can be defined via  anomalous dimensions of double- (or higher-) trace operators.  It is not surprising that in cases where one starts with a bulk theory, one can see features of the {\it reduced transition matrix elements} in the correlators and anomalous dimensions -- such as their growth with the coupling, {\it etc.} -- as is found in \GGP\ and \Katzetal. 
However, this does not yet determine the fine-grained S-matrix between appropriate scattering states; indeed from the bulk perspective, operator dimensions in the CFT are expected to receive contributions from the multiple scatters in the bulk.

\newsec{Boundary compact wavepackets}

To avoid the ``multiple-collision'' problem, one can consider particles incident from the boundary, as was done by Polchinski\PolS\ and Susskind\SussS.  As noted, a generic boundary source will produce a non-normalizable wavepacket, with the potential for interactions outside of our chosen interaction region.  We will discuss these questions in greater detail below, but we may initially try to avoid them by restricting the support of the boundary sources to small regions.

 Specifically, \refs{\GGP} proposed using such ``boundary compact'' wavepackets.  Consider a source of the form
\eqn\bcsource{f(b) = L(\tau, \ehat) e^{\pm i \omega R \tau}\ ,}
where $L(\tau, \ehat)$ is sharply peaked near a point $b_0=(\tau_0,\ehat_0)$, with widths $\delta\tau$ and $\delta \theta$, and has compact support, and $\omega$ gives the typical energy of the wavepacket seen by a central observer at $\rho\ll 1$.  
For such narrowly-peaked sources (compact support or not), a simple, explicit expression can be found for the corresponding wavepackets in the flat region $\rho\ll1$, generalizing \refs{\GaGi}.    
This follows from a Schwinger-like expression for the bulk-boundary propagator\GaGi,
\eqn\gbdelr{G_{B\partial}(b',x) = { (\cos\rho)^{\Delta}{\hat N} \over R^{(d-1)/2} }\int_0^\infty d\alpha \alpha^{\Delta-1} \exp\left\{i\alpha [\cos(|\tau-\tau'|-i\epsilon) - \sin\rho\, \ehat\cdot\ehat']\right\}\ ,}
where 
\eqn\nhatdef{{\hat N} = {2\Delta-d\over i^{\Delta-1} 2^{\Delta+1} \pi^{d/2} \Gamma(\Delta+1-d/2)}\ .}

In order to describe the wavepacket in the interaction region of size $\ll R$, one uses a choice of coordinates that become local flat coordinates in that region.   One choice, used in \FSS\ and \GaGi, is 
\eqn\firstcoord{t'=R\tau\ ,\ r'=R\rho\ .}
However, we find that we can reduce the deviations from the desired flat-space wavepackets\foot{This can be seen by comparing the expansions in $\delta\tau$, in the two coordinate systems.}  with a different choice\refs{\forty},
\eqn\barcoords{t = -R \cos(\tau -\tau_0)\ ,\ r = R \sin\rho}
 in which the metric takes the form
\eqn\rtmet{ds^2 = {1\over 1-r^2/R^2} \left(-{dt^2\over 1-t^2/R^2}+ {dr^2\over 1-r^2/R^2} + r^2 d\Omega^2_{d-1}\right) }
 This is a good approximation to the Minkowski metric for $t,r\ll R$. 
 
The bulk wavepacket  arising from \bdywavepack\ and \bcsource\ can then be expressed as a superposition of plane waves in the flat coordinates; we moreover set $\alpha = kR$ to write an expression\forty
\eqn\flatpack{\psi_f = \int_0^\infty {k^{\Delta-1}dk}\int d\ehat'\ F(k,\ehat', t,r) e^{-ikt-ik\ehat'\cdot {\vec r}}\ .}
Here,
\eqn\Fexact{F(k,\ehat',t,r) = {{\hat N}\left(1-r^2/R^2\right)^{\Delta/2} \over R^{(d-1)/2-\Delta}} \int d\tau' f(b') e^{ikR\left[\cos(|\tau-\tau'|-i\epsilon)-\cos(\tau-\tau_0)\right]}\ .}

In the case of narrowly-supported $f$, we can expand the exponent of \gbdelr\ about $\tau'=\tau_0$, which gives
\eqn\Fdef{F(k,\ehat',t,r) = {{\hat N}\left(1-r^2/R^2\right)^{\Delta/2} \over R^{(d-1)/2-\Delta}} \int d\tau e^{ik\sqrt{R^2-t^2}(\tau-\tau_0+i\epsilon)} f(\tau,\ehat')\left\{1+\calo[kt(\tau-\tau_0)^2]\right\}\ .}
For $t/R\ll1$, $r/R\ll1$, this simplifies further to a momentum-space shape
\eqn\flatpack{F(\veck) \approx {{\hat N} \over R^{(d+1)/2-\Delta}} \int d{\tilde t} e^{ik{\tilde t}} f(\tau_0 + {{\tilde t}\over R}, -{\hat k})\ ,}
independent of $t$ and $r$: the bulk momentum-space wavefunction is the Fourier transform of the boundary source, with time conjugate to magnitude of the momentum.

Such wavepackets do not have the usual analytic behavior in the momentum $k$.  Specifically, a regular function in $\veck$ has behavior $\sim k^{l}Y_{l\vecm}({\hat k})$ at $k=0$, and a common stronger requirement, for a ``regular wavepacket''\ReSi, moreover requires vanishing support in a neighborhood of $k=0$.\foot{In order to obtain a complete space of wavepacket states that are regular at $\veck=0$, one might try boundary $f$'s of the form $(\partial_\tau)^l S(\tau) Y_{l\vecm}(\ehat)$, where the $S$ are suitably-chosen compact and $l$-times differentiable functions ({\it e.g.} as in \GGP).  We leave such explorations for further work.}  Alternatively, for massless particles, scattering theory can be based on wavepackets with compact support in position.  One generically does not achieve these behaviors with the compact $f$ we have described of the form \bcsource.  Indeed, while wavepackets may be arranged with characteristic widths 
\eqn\widths{\delta t\approx R\delta\tau\ ,\ \delta x_\perp \sim 1/\delta p_\perp\sim 1/(\omega \delta \theta)\ ,}
outside these ranges
\GaGi\ argues that such wavepackets fall off only as a power law,
\eqn\psiasymp{\psi_f \sim 1/r^\Delta\ ,}
in both longitudinal and transverse directions.

An obvious concern is that such tails can make unwanted contributions to scattering behavior.  As an extreme example, consider the collision of two wavepackets with ultraplanckian center-of-mass energy.  Even with the tails, two such wavepackets can be well-enough focused in impact parameter to form a black hole with significant probability.  Specifically, the energy $\omega$ can be increased, at fixed $\delta t$ and $\delta x_\perp$, so that the Schwarzschild radius  $R_S(\omega)$ is much larger than these widths.  However, there remains some probability for particles in the tail of the wavepackets to scatter at impact parameters larger than $R_S(\omega)$, avoiding black hole formation.  In contrast, with Schwartz wavepackets, the latter amplitudes can be made exponentially small in a power of the scattering energy.

This appears to be relevant for correct description of S-matrix elements in the black hole regime.  It has been argued (for further discussion in this context see \refs{\GiSr,\GiPo,\Erice}) that the amplitudes to specific final states of Hawking radiation from such a black hole are of order $\exp\{-S/2\}$, where $S\sim E^{(D-2)/(D-3)}$ is the Bekenstein-Hawking entropy.  The ``meat'' of the very important problem of information in black hole formation/evaporation then lies in understanding the origin of precise relations between these amplitudes needed for unitarity.   In particular, the two-to-two amplitude is expected to be exponentially small in $S$.  But, with power-law tails, one can have a larger contribution to the amplitude from the particles that ``miss'' the black hole region.  This is an example of potential problems of resolution arising from such tails; ref.~\GaGi\ also discusses how such tails can obscure Rutherford-scattering behavior.

A related question is the origin, from the boundary perspective, of various detailed scattering behavior expected for the bulk S-matrix.  First of all, the bulk S-matrix should conserve momentum (up to $\sim 1/R$ corrections); \GGP\ found that this can arise, in a plane-wave limit for wavepackets, if the boundary correlators have a certain kind of singularity structure whose origin is rather mysterious from the perspective of the boundary theory.  The S-matrix should also have other features reviewed in \Erice; for example, in the case of ultraplanckian scattering, at large impact parameters the scattering should be well approximated by the Born amplitude $\sim G s^2/t$, and then at shorter distances one expects onset of eikonal and then strong gravity/black hole behavior.  From the boundary perspective, it is not at all clear what could be responsible for these transitions in behavior.  

One way to approach this problem is to investigate the boundary appearance of the scattering wavepackets \flatpack.  Extending arguments in \GaGi, we find that in the state created by \bcsource, the corresponding boundary operator has a correlator
\eqn\ocorr{\langle0| \calo(b)|\psi_f\rangle \approx (2\Delta-d)R^{(d-1)/2}  \int d^dk k^{\Delta-d} F(\veck) e^{iR[k\cos(\tau-\tau_0) - \veck \cdot \ehat]}\ .}
For a sharp boundary source, this boundary signal (which is linked to the bulk tails) propagates as a ring, first expanding and then contracting, around the  boundary.  In the collision of two such signals, the question then becomes that of how adjustment of the boundary wavepacket corresponding to changing the bulk impact parameter leads to the scattering transitions described above.

Of course, we would prefer to examine this dependence for more familiar scattering states.  It is instructive to  
understand the kind of boundary source needed to produce such a state.  As an example, consider a gaussian,
\eqn\gausspack{F(\veck)= e^{-i \veck\cdot \vecx_0} e^{-{(\veck-\veck_0)^2\over 2 (\Delta k)^2}}\ .}
Inversion of the Fourier transform in \flatpack\ gives
\eqn\gaussource{f(\tau,\ehat) = {R^{(d+1)/2-\Delta}\Delta k\over\sqrt{2\pi}{\hat N}} \exp\left\{ -{(\Delta k)^2\over 2} \left[R(\tau-\tau_0) - \ehat\cdot \vecx_0 - {i\veck_0\cdot \ehat\over (\Delta k)^2}\right]^2 - {k_0^2\over 2(\Delta k)^2 }\right\}\ .}
Note in particular that a shift of impact parameter corresponds to $\vecx_0\rightarrow \vecx_0+{\vec b}$, with ${\vec b}\perp \veck_0$.
However, the source \gaussource\ is not well-localized on the boundary, and if one wishes to use such sources one needs to address the questions that they pose.  Compact support wavepackets are even more puzzling, as under the map \boundcorr\ they have vanishing signal on the boundary\GaGi.

\newsec{Issues for non-normalizable states}

As described above, at a time $\tau$ where a boundary source $f(b)$ has support, the corresponding bulk field has non-normalizable behavior, \psinn, which as we have explained can be thought of as describing an infinite particle number.  Thus, sources without compact support generically do not create states in the original bulk Hilbert space, and this raises questions about how such sources can be used.  
  The first potential problem is that of unwanted interactions, due to the large integrated amplitude in the infinite regions near the boundary; a second is that of how to formulate a physical unitarity condition in the bulk.

To see the question of interactions, consider contributions to a scattering amplitude with nonnormalizable sources.  Following the rules of \refs{\GKP,\WittAdS} (which one might derive from an interaction-picture treatment like in section three), and considering for example a trilinear scalar interaction, we expect for example to have expressions of the form
\eqn\divexp{\int dV \, \psi_{NN} \psi_{NN} G_{Bulk}\ ,}
where a bulk Green function, with normalizable behavior, is integrated against a pair of non-normalizable wavefunctions.  This can produce a divergence near the boundary, which corresponds to infinite interaction amplitude there.

However, the gravitational interaction is special, since its effective coupling is the energy.  More exactly, the (super)graviton couples through the (super)stress tensor. For a state with a fixed bulk energy $\omega$ at the center of AdS, like those discussed in the preceding section, the proper stress tensor seen by observers near the boundary redshifts to zero as the boundary is approached.  This is easily seen by writing the metric \glocoora\ in terms of a vielbein $e^a_\mu$, giving the local proper stress tensor
\eqn\properstress{T_{ab} = e_a^\mu e_b^\nu T_{\mu\nu}\ .}
This results in an extra factor of $\cos^2\rho$ in the proper stress tensor, due to this red shift.

One also needs the behavior of the graviton propagator. This is gauge dependent, but one can bound its falloff at the AdS boundary by finding a gauge in which it exhibits a particular falloff.  (Exhibiting a particular falloff does not guarantee that there is no gauge with faster falloff.)  Specifically, we can use the expressions derived in \DHFr.   There, the bi-tensor structure of the propagator  under the isometry group  $SO(d,2)$ of AdS$_{d+1}$ was used to decompose it into invariant components; \DHFr\ treats the case of AdS${}_5$.  This produced five terms, which are products of an invariant bi-tensor $\calo^{(i)}$ and a scalar function of the geodetic distance $u$,  $G_i(u)$. A convenient gauge for the metric perturbation $\delta g_{\mu\nu}=h_{\mu\nu}$ is the ``Landau gauge,''
\eqn\LandauGauge{\nabla^\mu h_{\mu\nu}={1\over d+1}\nabla_\nu h\ .}

The authors of \DHFr\ found  algebraic relations amongst the functions $G_i$ by imposing the gauge conditions, which reduce the problem to solving a single ODE. From their construction one finds that all of the functions $G_i$ fall as $u^{-4}$ at large $u$, which corresponds to one of the endpoints approaching the boundary. This follows once one corrects their formula (5.48), which should read
\eqn\GCorrect{
  G_2(x) = -4(x-1)^2\left({x^2\over n(n+1)}g''(x) + {2x\over n}g'(x) + g(x)\right)\ ,
}
where $x={u\over2}+1$.  This correction is needed for the propagator to satisfy the gauge condition, and also to be consistent with 
with the Fefferman-Graham form of the metric in an asymptotically AdS space \FeffGr
\eqn\feffGraham{
ds^2 = {1\over\cos^2\rho}(-d\tau^2+d\rho^2+\sin^2\rho d\Omega^2)\left[1 + \calo(\cos^4\rho)\right]\ .
}
The falloff behavior $1/u^4$ corresponds to the behavior 
\eqn\metfall{\langle h_{ab} h_{cd}\rangle \sim \cos^4\rho}
of the metric propagator components in a local frame, as one of the points approaches the boundary.  Note that this behavior is the same as the falloff for a massless scalar field.

Combining these results,  the $\rho$ integral in \divexp\ for gravity in AdS${}_5$ now behaves asymptotically as
\eqn\convexp{\sim \int {d\rho \over \cos^5 \rho} (\cos \rho)^{2(d-\Delta)} \cos^6\rho\ ,}
and in particular is convergent for the massless case $d=\Delta$.
Even though this discussion indicates that interactions are not {\it infinite} (for small enough mass) due to non-normalizable behavior, in any specific proposal for construction of states with non-compact sources one must also check that the finite interactions of particles outside of the interaction region are not {\it significant}.  

As noted, another potential issue with non-normalizable solutions is that of formulating a unitarity condition for the bulk S-matrix.  The statement $S^\dagger S=1$ is well-formulated on a Hilbert space of states; for example one may pick a basis of such states, and write it as a matrix relation, $S_I^{\dagger J} S_J^K=\delta_I^K$.  But, it is less clear how to formulate it on states or wavefunctions without a finite inner product.

Before returning to the latter question in section seven, it is instructive to try to take advantage of the finiteness \convexp\ of gravitational interactions for non-normalizable wavefunctions to find boundary sources with weaker restrictions on their support, but which produce states that are normalizable after a given time.  Study of such an intermediate construction sheds more light on the restrictions on construction of scattering states.

\newsec{Resonant wavepackets}

To avoid the complications of non-compact boundary sources, which do not produce normalizable states, section four considered sources with compact support.  However, these were found to have power law tails.  The preceding discussion suggests, however, that we might relax restrictions on the boundary sources.  To avoid questions of normalization, we will assume that the source turns off and we have a normalizable ``in'' state after a specific time, and in particular in the interaction region, but we might allow source support that is non-compact (or non-compact in the $R\rightarrow\infty$ limit) prior to this time.  (For ``out'' states, we reverse the time.)   In particular, we will find that we can construct an {\it arbitrary} normalizable wavepacket \gennorm, after a given time taken without loss of generality to be $\tau=0$.  
Eq.~\convexp\ ensures that such overlapping sources do not lead to divergent interactions, but we do need to check for possible relevance of the resulting finite interactions.

Specifically,  consider boundary data $f(b)$ with support only for $\tau\leq0$. It is possible to take advantage of the resonant structure of AdS to build such a wavepacket in half an AdS period.\foot{Or, if desired, one may use more such intervals.} Let us write $f(b)$ in terms of modes on the boundary, 
\eqn\boundSource{
  f(b)=\left[\theta(\tau+\pi)-\theta(\tau)\right]\sum_{nl\vecm}f_{nl\vecm}Y_{l\vecm}(\ehat)e^{-i\omega_{nl}\tau}\ .
}
The bulk wavefunction $\psi_f$ is then given by \bdywavepack,
\eqn\bulkWavefn{\eqalign{
  \psi_f(x)&=\int db' f(b') G_{B\partial}(b',x) \cr
  &= (2\Delta-d)R^{(d-1)/2} \int_{-\pi}^0d\tau' \int d\ehat' f(b') \int{d\omega \over 2\pi} \sum_{nl\vecm} {k_{nl}Y_{l\vecm}(\ehat')^*\phi_{nl\vecm}(\vecx)e^{i\omega(\tau'-\tau)} \over \omega_{nl}^2-\omega^2-i\epsilon}\cr
  &= i(\Delta-{d\over2})R^{(d-1)/2}\sum_{nn'l\vecm}\int_{-\pi}^0d\tau'{f_{n'l\vecm}k_{nl}\phi_{nl\vecm}(\vecx)e^{-i\omega_{n'l}\tau'}\over\omega_{nl}}\Bigl[\theta(\tau'-\tau)e^{-i\omega_{nl}(\tau'-\tau)}\cr &\quad\quad\quad\quad\quad\quad\quad\quad\quad\quad\quad\quad\quad\quad\quad\quad\quad+\theta(\tau-\tau')e^{i\omega_{nl}(\tau'-\tau)}\Bigr] \ ,
}}
where we use the mode-expansion \FSS\ for the bulk boundary propagator, and
\eqn\knldef{k_{nl}= {e^{\pm i \pi n}\over \Gamma(\Delta+1-d/2)} \sqrt{2\onl\over R^{d-1}}\sqrt{\Gamma(\Delta+n+l)\Gamma(\Delta+n+1-d/2)\over\Gamma(n+l+d/2)\Gamma(n+1)} \
.}

For $\tau>0\geq\tau'$, only the second term in \bulkWavefn\  contributes. Making use of the fact that
\eqn\KronDeltaInt{
  \int_{-\pi}^0 d\tau' e^{i(\omega_{nl}-\omega_{n'l})\tau'} = \int_{-\pi}^0 d\tau' e^{2i\tau'(n-n')} = \pi \delta_{nn'}\ ,
}
we find
\eqn\WavefnPosTime{
  \psi_f(\vecx,\tau'>0) = \sum_{nl\vecm}2\pi i(\Delta-{d\over2})R^{(d-1)/2}{f_{nlm}k_{nl}\over\sqrt{2\omega_{nl}}}\phi_{nlm}(x)\ .
}
Thus, if we take
\eqn\sourceGenNorm{
  f_{nl\vecm}={c_{nl\vecm}\sqrt{2\omega_{nl}} \over 2\pi i(2\Delta-d)R^{(d-1)/2}k_{nl}} \ ,
}
we can indeed produce an arbitrary normalizable state \gennorm\  for $\tau>0$. 

Note that in the limit $R\rightarrow\infty$, the boundary source \boundSource\  becomes non-compact in bulk physical time $t=R\tau$. This behavior, which is not present in the earlier boundary-compact wavepackets of \refs{\GGP,\GaGi}, allows the construction of arbitrary, and in particular, regular, wavepackets in the interaction region. 

Note also that if we push $f(b)$ infinitely far into the past by shifting $\tau\rightarrow\tau-T$, $T\rightarrow-\infty$, then we have an explicit construction of the free ($g=0$) operator-state correspondence, in which normalizable states in the bulk correspond to states in the CFT by the AdS/CFT dictionary. From the boundary perspective, this is clear, since after a conformal transformation from the cylinder to the plane, shifting the source $f(b)$ into the infinite past maps to localizing the operator at the origin in radial quantization. 

 \Ifig{\Fig\respack}{A resonant wavepacket is built up by a boundary source acting during the half-period indicated by the dark bars.  During this time, as is described in the text, the wavepacket grows in amplitude to the final value.}{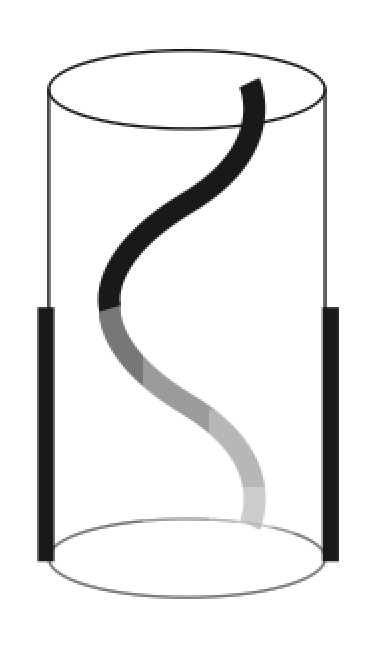}{1.5}

While the resonant construction gives arbitrary normalizable wavepackets for $\tau>0$, and while the results of the preceding section ensure that interactions of two such wavepackets will not be infinite for $\tau<0$, we still need to check whether the interactions are significant.  Specifically, the source is turned on for a long time to build the wavepackets, so if the wavepacket is built slowly over time, there is a possibility of an analog to the multiple scattering problem outlined in section three.   For an illustration of some features of resonant wavepackets, see \respack.

First consider the behavior in the far past, before any sources are turned on. In this region the bulk wavefunction is purely negative frequency due to the causal properties of the Feynman bulk-boundary propagator. Thus, any interactions before the sources are turned on do not contribute to the physical scattering amplitudes. Furthermore, in the case of resonant wavepackets, it is clear from the form of \bulkWavefn\ that if $\Delta$ is an integer, $\psi_f$ vanishes identically. 

However, once the boundary source is turned on, both positive and negative frequency modes are present; then there can be secondary interactions contributing to the total amplitude, hindering our attempts at isolating a single, well-localized scattering event. Of particular concern is the possibility that the wavefunction will have significant support, and comparable to the size of the final wavepacket, at an ``image'' interaction region, seperated by a half-integer number of AdS periods from our desired interaction region.  As we have noted, in the free, sourceless theory these regions are of particular concern because the wavefunction is exactly periodic.

For example, in the case of the resonant wavepackets described by \bulkWavefn, we might worry that, if  we choose the time of the interaction region to be $\tau\approx \pi/2$,
at the image $\tau\approx-\pi/2$, $\psi_f$ will be of the same size as it is at the interaction time.    This would generically produce such secondary  contributions to the total scattering amplitude.

We examine this case in detail; while the source is turned on, $-\pi<\tau<0$, we find
\eqn\wavefnReflected{\eqalign{
  \psi_f = \sum_{nn'l\vecm}{c_{n'lm}\over \pi\sqrt{2\omega_{n'l}}} {\omega_{n'l}k_{nl} \over \omega_{nl}k_{n'l}}\phi_{nl\vecm}(\vec x) \Biggl[& e^{-i\onl\tau}\int_{-\pi}^\tau d\tau' e^{i(\onl-\omega_{n'l})\tau'}\cr&+ e^{i\onl\tau}\int_\tau^0 d\tau' e^{-i(\onl+\omega_{n'l})\tau'}\Biggr]\ .}}
%
%
%
By considering the cases $n=n'$ and $n\neq n'$ separately, this solution for $-\pi<\tau<0$ can be written as a sum of three terms, $\psi_f =\psi_f^{(A)} + \psi_f^{(B)} + \psi_f^{(C)}$, where
\eqn\reflectedPacket{\eqalign{ 
  \psi_f^{(A)} &= \left(1+{\tau\over\pi}\right) \sum_{nl\vecm}c_{nl\vecm}\phi_{nl\vecm}(x)\ ,\cr
  \psi_f^{(B)} &=  \sum_{nl\vecm}{c_{nl\vecm}\over 2\pi i \onl} \left[ \phi_{nl\vecm}(x) -\phi_{nl\vecm}^*(x)\right]\ ,\ {\rm and} \cr
  \psi_f^{(C)} &= \sum_{n\neq n',l\vecm}{c_{n'l\vecm}\over i\pi \sqrt{2\omega_{n'l}}} {\omega_{n'l}k_{nl}\over \omega_{nl}k_{n'l}}{\phi_{nl\vecm}(\vecx)}\left[{e^{-i\omega_{n'l}\tau}-e^{-i\omega_{nl}\tau}\over\omega_{nl}-\omega_{n'l}}-{e^{i\omega_{nl}\tau}-e^{-i\omega_{n'l}\tau}\over\omega_{nl}+\omega_{n'l}}\right]\ .
}}
The first contribution $\psi_f^{(A)}$ is exactly the kind of term that concerned us. Specifically, we see that the wavepacket builds linearly in time starting from when the source is turned on. Thus, at time $\tau=-\pi/2$, the term $\psi_f^{(A)}$ will be half the magnitude of $\psi_f(\vecx,\tau=\pi/2)$, and is equally-well localized as a wavepacket in the center of AdS. If we try to build two wavepackets of this form to scatter at time $\tau=\pi/2$, then we will find a contribution to the amplitude of the same magnitude coming from this reflected scattering in an image interaction region.

Of course, this presumes that  $\psi_f^{(B)}$ and $\psi_f^{(C)}$ do not destructively interfere with $\psi_f^{(A)}$ and eliminate the problem of secondary scattering. We can check this, in the large-$R$, fixed-$\omega$ limit relevant for flat-space physics, 
\eqn\continuum{
\omega_{nl}\rightarrow \omega R\quad ;\quad \sum_n\rightarrow\int_0^\infty R d\omega/2\ .}

Indeed, the second term $\psi_f^{(B)}$ vanishes in this limit. Turning our attention to the third term, consider specifically $r=0$, which projects on $l=0$.  Using the asymptotics of $k_{nl}$ (see \FSS, eq.~(A.47)) and performing the integrals over $\omega$ gives
\eqn\ThirdTerm{\eqalign{\psi_f^{(C)}(\vecx\approx0,\tau) \approx \left({1\over 2}\right)^{d-2\over 2}{\Gamma(\Delta)\over \Gamma(d/2)} \sum_n& {c_{n00}\over i\pi \sqrt{2\omega_{n0}}}  \left({\onz\over R}\right)^{d-1\over 2} e^{-i\onz(\tau+\pi)}\cr &\left[\Gamma(1-\Delta, -i\pi\onz/2)-\Gamma(1-\Delta,-i\onz(\tau+\pi/2))\right]\ .  }}
Then, using  \continuum\ again and the asymptotics of the incomplete gamma function gives
\eqn\thirdasymp{\eqalign{ \psi_f^{(C)} (\vecx\approx0,\tau) \approx \left({1\over 2}\right)^{d-2\over 2}{\Gamma(\Delta)\over \Gamma(d/2)} \sum_n& {c_{n00}\over i\pi \sqrt{2\omega_{n0}}} \omega^{d-1\over 2} e^{-i\omega(t+\pi R)}\cr & \left\{ {e^{i\pi R\omega/2}\over [-i\pi R\omega/2]^\Delta} - {e^{i\omega(t+ \pi R/2)}\over [-i\omega(t+ \pi R/2)]^\Delta}\right\}\ .}}
%
This has additional power-law falloff $\sim1/[\omega(t+\pi R/2)]^\Delta$ in time, so is generically not of a form to destructively interfere with the reflected wavepacket $\psi_f^{(A)}$.  Moreover, the resulting non-normalizable behavior introduces the possibility of further interactions, for example between the $\psi^{(C)}$ components of two waves.  A non-trivial (and so-far unexplainable) cancellation would be required between the different scattering contributions to keep their effect from competing with scattering in the interaction region.

One might be concerned that the magnitude of $\psi_f$ is infinte at $\vecx=0,\tau=-\pi/2$. This divergence is an artifact due to the sharp turn-on/off of the source at times $\tau=-\pi,0$, which could be smoothed out by smoothing the theta functions in \boundSource.  Of course, this then complicates the problem of precisely constructing a desired arbitrary bulk state.

Indeed, one could in fact consider more general behavior than our specific resonant wavepackets, where the source is for example turned on more gradually, or even over infinite time.  However, our construction exhibits what appears to be general features: in addition to the non-normalizable fall-off, such sources will also produce a wavepacket in images of the interaction region, that is of size comparable to the final wavepacket.  Extending the source over longer times only gives more opportunity for secondary interactions, contaminating the amplitude for the primary event in the interaction region.

\newsec{The question of unitarity}

Turning to another general consideration, the common statement that unitarity in the boundary theory ensures unitarity of the bulk theory is one worth further examination, particularly given hopes that this would play a role in resolving the unitarity crisis (or black hole information ``paradox'') in gravitational scattering.  

First, the meaning of unitarity in the boundary theory is clear:  if it is a field theory with a hamiltonian, as with ${\cal N}=4$ super Yang-Mills, there is an evolution operator $U=\exp\{-i\tau H\}$ satisfying
\eqn\boundunit{U^\dagger U=1\ .}
Then, if there is a unitary map \hilbmap\ between Hilbert spaces, and such that flat-space scattering states and an S-matrix can be extracted in a controlled approximation, this should obey bulk unitarity
\eqn\bulkunit{S^\dagger S=1}
in that approximation.  

As we have stated, for non-compact sources, which produce non-normalizable bulk configurations we might refer to as ``pseudostates,'' we do not have a map \hilbmap.  Moreover, the non-normalizablity hinders formulation of the unitarity condition \bulkunit.  For example, in a theory with a small expansion parameter, we commonly decompose 
\eqn\Tmat{S=1+i{\cal T}\ , }
where $\cal T$ vanishes in the limit of small coupling.  In a Hilbert-space context, the unitarity condition \bulkunit\ at zeroth order in the coupling is then the statement that $1=\sum_I|I\rangle\langle I|$ squares to one, via the orthonormality condition $\langle I|J\rangle = \delta_{IJ}$.  We do not have this structure without a space of normalizable states.  While one could attempt to regulate the non-normalizable behavior through a UV cutoff in the boundary theory, it is not clear how to do so in a way that is intrinsically formulated in that theory, yet yields sensibly-regulated physics from the bulk perspective.\foot{One could also consider the option of regulating using boundary counterterms as in \refs{\AnMa}, but this type of regulation was argued there to lead to boundary ghosts.}

If there is a construction of appropriate asymptotic states that gives a good approximation to a physical bulk S-matrix, an interesting question regards the possible origin of the bulk unitarity condition \bulkunit.
Schematically, as we have indicated, the suggested correspondence is between certain correlators (perhaps involving integrals of boundary operators against particular sources) and S-matrix elements, {\it e.g.}
\eqn\schemcorr{ \langle \calo_1\cdots \calo_n\rangle \ \leftrightarrow \ S[\psi_1,\cdots,\psi_n]\ .}
A possible analog to the decomposition \Tmat\ is, taking for example the four-point function,
\eqn\disconn{\langle \calo_1\cdots \calo_4\rangle = \langle \calo_1\calo_2\rangle \langle \calo_3\calo_4\rangle + \langle \calo_1\calo_3\rangle \langle \calo_2\calo_4\rangle+ \langle \calo_1\calo_4\rangle \langle \calo_2\calo_3\rangle +\langle\calo_1\cdots \calo_4\rangle_c\ ,}
isolating the connected piece of the correlator\forty.\foot{We thank T. Okuda and J. Polchinski for conversations on this point; \FiKa\ has independently utilized such a subtraction.  Note that, for noncompact sources one would also need an appropriate means to regulate this expression, since the individual terms are infinite.}  

 \Ifig{\Fig\onepf}{Factorization on an intermediate single-particle state.}{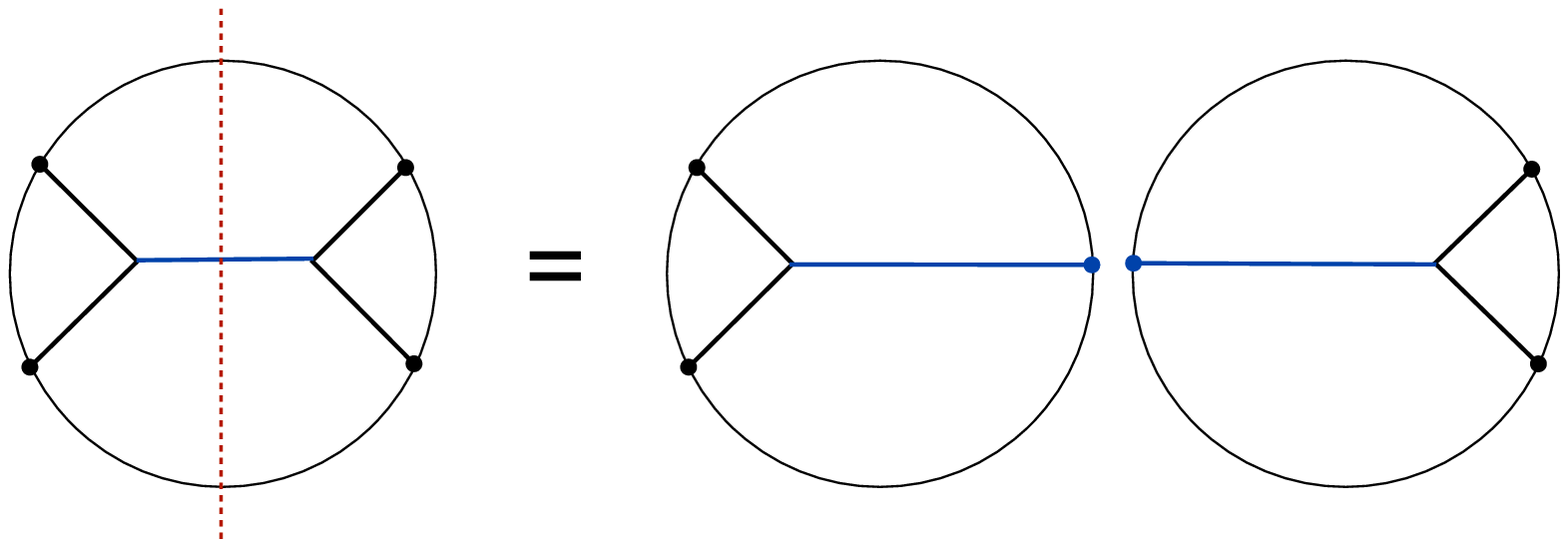}{4.5}
 
 \Ifig{\Fig\twopf}{Schematic of a possible factorization relation representing   a two-particle cut in the bulk.}{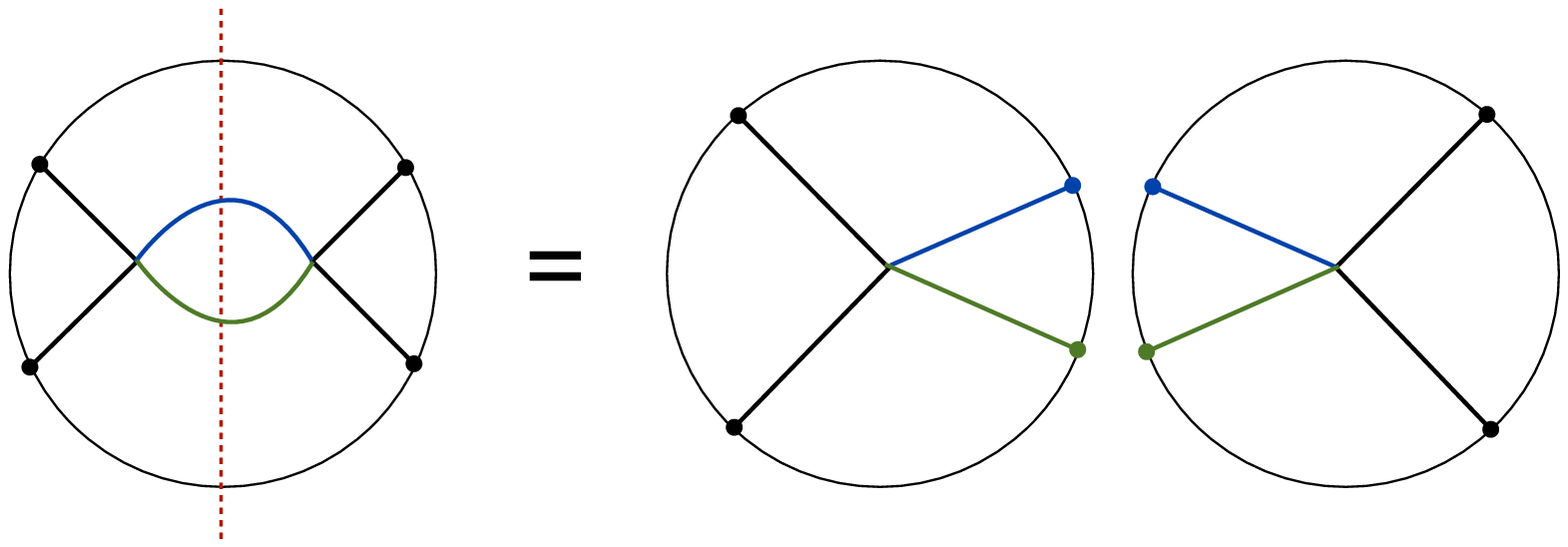}{4.5}

If so, the bulk unitarity condition \bulkunit, which can be restated as
\eqn\unitreform{i({\cal T}^\dagger - {\cal T}) = {\cal T}^\dagger {\cal T}}
should have an equivalent boundary statement that holds, at least in a good approximation, for these correlators.  In the case of a single-particle intermediate state, say in the s-channel, as pictured in \onepf, an apparent analog can arise from the OPE: 
\eqn\OPE{\langle\calo_1\cdots \calo_4\rangle_c \sim \sum_k \langle\calo_1 \calo_2\calo_k\rangle \langle\calo_k\calo_3 \calo_4 \rangle_c\ .}
Indeed, related unitarity constraints on interactions were described in \refs{\Katzetal,\Raju}.
But, familiar bulk S-matrices also have cuts corresponding to multiparticle states.  For example, in the case of a two-particle intermediate state, an interesting question is how a two particle cut, like that pictured in \twopf, would arise.  Written in terms of boundary correlators, this is suggestive of a relation of the form
\eqn\twocut{2{\rm Re}\langle\calo_1\cdots \calo_4\rangle_c \sim \sum_{k,l} \langle\calo_1 \calo_2\calo_k \calo_l\rangle_c \langle\calo_k \calo_l\calo_3 \calo_4 \rangle_c\ }
expressing the correct analytic structure.
A very interesting question is how such unitarity relations, and other multiparticle generalizations, could  originate in an appropriate boundary theory. (An intermediate possibility is that such a unitarity condition can be formulated on a space of of ``coarse-grained'' states  -- like for example that of our boundary-compact states with tails -- but not on fine-grained bulk states.)

\newsec{Summary and conclusions}

In order for AdS/CFT to give a non-perturbative definition of string theory, we need a prescription to compute directly in the boundary CFT physical quantities describing bulk physics.  Since the bulk theory is gravitational, the most natural target is an approximation to the flat-space S-matrix.   As  one might expect, extraction of a higher-dimensional S-matrix with the correct properties from a lower-dimensional theory is not trivial.  Let us summarize constraints on the particular proposal of AdS/CFT.  Description of an S-matrix requires defining appropriate scattering states, and the amplitudes for transitions between them.  As we have seen, there are non-trivial issues with both.  

In AdS, wavepackets built from normalizable states do not asymptotically separate, and indeed in the free theory are periodic up to an overall phase.  This hinders a description of scattering where there is a controllably small interaction that only acts in a given ``interaction region;'' instead, amplitudes involving such states will generically receive contributions from scattering in an infinite number of mirror interaction regions.  

One may try to avoid these infinite interactions by producing wavepackets via sources at the boundary.  A generic boundary source will not produce a normalizable state, raising both questions of how to correctly formulate unitarity of the bulk S-matrix, and again the possibility of unwanted interactions outside the interaction region contaminating amplitudes.  One approach to avoiding these difficulties is to use boundary sources that are compact in the large-$R$ limit\refs{\GGP}.  But, as we have described here and in \refs{\GaGi,\forty}, such sources do not appear to produce the kinds of wavepackets used in careful treatments of scattering theory, and in particular have power-law tails which limit  resolution; these can be thought of as arising from extra contributions near zero momentum, that don't exhibit the usual momentum-space regularity.

We have also found a construction intermediate between these ``boundary-compact'' wavepackets and normalizable states, the ``resonant'' wavepackets of section six.  While these eliminate the unwanted tails, they require a time of order $R$ to build up the wavepacket, where they can generically interact outside the interaction region, particularly in an image interaction region.  In the absence of non-trivial cancellations, this generically indicates scattering amplitudes of comparable size to those for scattering in the interaction region, which interfere with our attempts to isolate the latter.
This construction also illustrates crossover between compact-support sources with tails and normalizable wavepackets with multiple interactions.

In a similar construction, \refs{\FiKa} have proposed use of rapidly-falling non-compact boundary sources -- see their eqns.~(2.9) or   (2.20).  If the wavefunction is interpreted to arise from the integral of such a function against a boundary Heisenberg picture operator in the interacting theory, passage to the interaction picture as in section three then gives an integral of the source against a Feynman bulk-boundary propagator, like in sections four and six, and thus produces a bulk non-normalizable wavefunction.  Depending on further details of the construction, these could either have tails as described in section four, or a multiple interaction problem such as in section six; to be viable one would need to explicitly check that neither problem is present.\foot{If, on the other hand, one interprets formulas of \FiKa\ to give an integral of the 
the Wightman version of the bulk-boundary two-point function against such a source (though this doesn't naturally appear in the GKPW calculation of boundary correlators) one produces a bulk wavefunction that is periodic, and thus encounters the multiple-interaction problem.}  They in addition face the questions of normalization and unitarity outlined in sections five and seven.

Even if there were a suitable definition of scattering states, the need to recover physical bulk amplitudes would require other non-trivial structure in the boundary theory.  For example, \GGP\ showed that in order to recover (approximate) bulk momentum conservation, the boundary theory must have certain singularities when the boundary points of a correlator are all lightlike separated from a common bulk point.  Moreover, in order to reproduce expected reduced transition matrix elements -- for example $T\sim G s^2/t$ in the Born regime, these correlators must have certain subleading singular structure.  Ref.~\refs{\HPPS} gave some evidence for the plausibility of these structures, but how and why they would emerge in appropriate boundary theories is at best incompletely understood.  

A bulk S-matrix has other crucial properties.  One is unitarity, as described in section seven.  The origin of the unitarity conditions \bulkunit\ or \unitreform, which we might expect to be recast in ``cutting" relations like \twocut, is unclear.  Moreover, while we would expect such a boundary construction not to describe a precisely local bulk theory, it should have sufficient approximate bulk locality to give a viable description of the kinds of phenomena we observe.  Without a notion of local observables, there is no direct formulation of microcausality (commutativity of local observables outside the light cone).  But, other notions of locality for the S-matrix include clustering and macrocausality;\foot{For further discussion, see \refs{\Coleman,\Erice}.} the latter states that the leading correction to large-distance clustering involves single-particle exchange between the two clusters.  A good question is how these would be obtained; they appear to require a boundary theory to satisfy further non-trivial conditions.  

Yet another challenge is that we have dealt with light states, with masses $m\roughly< 1/R$.  Generalization of our constructions  faces other questions  in the case $m\gg 1/R$ -- as would be relevant for a particle with mass of the electron in an AdS space with radius $10^{10}$\ light-year!  These include divergences in \divexp, and greater difficulty in forming desired wavepackets.

In short, while we have not proven a no-go theorem for a boundary theory to give a sufficiently close approximation to fine-grained bulk physics, the challenges appear formidable and possibly insurmountable.  
An alternative\refs{\FSS,\GaGi,\forty} is that while AdS/CFT apparently provides a correspondence in an appropriately coarse-grained sense, and one incidentally with very interesting implications for boundary physics, it does not furnish a fine-grained description of bulk physics.

\newsec{2016 update}

Although this was not part of the subject of this paper, the referees asked that we address the question of developments in the ``Mellin amplitude" program; we will also make some comments on other developments since the paper originally appeared.

First, for perspective, a main focus of this paper has been on the question of recovering the bulk S-matrix from the boundary theory, via a procedure that does not directly rely on bulk physics.  Recall that, in general, the S-matrix gives amplitudes for transitions between scattering states of a theory.  Thus two questions are those of defining appropriate scattering states, and then of showing that one has a procedure to calculate the correct amplitudes connecting them. The discussion of sections two through six is largely focussed on the first question, of identifying a boundary construction that yields scattering states with the properties expected from bulk physics; here we have examined a significant set of challenges, which still appear not to be resolved.

There was initially considerable hope expressed by advocates that the Mellin program gave a pathway to extracting the flat-space S-matrix.  However, after roughly five years of development, these do not appear to have been fully realized.  The basic Mellin approach (see {\it e.g.} \refs{\FitzpatrickIA}) writes CFT correlations functions in terms of a Mellin transform of the Mellin amplitude.  If this can be inverted to find the Mellin amplitude, then there is also a formula giving a corresponding reduced transition-matrix element in terms of the Mellin amplitude. There have been some limited results in applying this procedure to cases of interest, but there are questions about how, in general, one derives Mellin amplitudes from correlators; there is potential non-uniqueness in determining the Mellin amplitudes, and questions of how to treat higher-loops and massive particles remain\refs{\PenedonesS}.

In particular, while the mathematical structure of the Mellin amplitudes has some suggestive mathematical elements, it should be emphasized that as just noted, so far it, optimistically, just gives a prescription to extract reduced transition-matrix elements, not elements of the S-matrix.  The latter are matrix elements labelled by scattering states; the former are more like couplings of the theory.  That is, if one assumes one has a bulk S-matrix with certain structure, the Mellin approach potentially allows one to read off some of the couplings of the bulk theory.  This is not, yet, a construction of the full matrix structure of the S-matrix, and in particular does not appear to address the questions regarding identifying  correct scattering states, such that the S-matrix has the correct bulk structure.  

A related issue regards the question of unitarity conditions that we discuss; while refs.~\refs{\FitzpatrickHU,\FiKaUnitarity} have suggested a unitarity relation like eq. \twocut, where cuts arise via coalescence of poles from factorization on multi-trace operators, questions remain in particular related to the identification of the scattering states.

Since the original version of this paper appeared, there have been other developments
that serve to sharpen some of the issues we have put forward. 

For example, section three makes the point that we do not necessarily find a stable perturbative expansion in $g$, when considering normalizable states.  This appears to have been reinforced by results on the stability of AdS.  While flat space is stable, with small perturbations dissipating away to asymptotic infinity \refs{\ChristodoulouKlainerman}, AdS has been shown to be unstable once $g$ is nonzero and interactions are turned on.  This is due to the nature of the reflecting boundary conditions, which result in small perturbations in AdS focusing and forming black holes \refs{\AndersonAX\BizonRostworowski-\DiasSS}. This specifically appears to lend further support to our claims that in interacting theories in AdS, the infinite multiple scatterings of normalizable multi-particle states confound the separation of a single amplitude. Such multi-particle states are unstable to forming black holes in regions other than the single interaction region of interest.

More generally, the question of how and whether we can ``decode the hologram" remains important.  There are two basic approaches to this\forty: one is identifying operators in the CFT that correspond to some approximate version of local bulk operators; the other is to try to extract the bulk S-matrix from the CFT.  While there has been a lot of recent focus on the former approach, the question of the latter providing a ``decoding map" is one of ongoing interest.  Developments in black hole physics have also continued to emphasize our lack of understanding of AdS/CFT, even leading one previously staunch AdS/CFT advocate to state\refs{\Polcomm} that ``...the CFT really doesn't give you a good description of the bulk for many reasons, and you need an independent theory of the bulk."  Further clarification of the true status of AdS/CFT, from both the operator and S-matrix perspective, is needed.  This paper summarizes some apparently important questions regarding how and whether it is possible to construct the bulk S-matrix, particularly in the large-$R$ limit, from the boundary theory.

\bigskip\bigskip\centerline{{\bf Acknowledgments}}\nobreak

We wish to thank J. Kaplan, E. Katz, T. Okuda, and particularly D. Marolf and J. Polchinski, for many useful conversations.
This work  was supported in part by the Department of Energy under Contracts DE-FG02-91ER40618 and DE-SC0011702, by grants FQXi-RFP3-1008, FQXi-RFP-1507 from the Foundational Questions Institute (FQXi)/Silicon Valley Community Foundation, and by the FWF project P 27396-N27.

\listrefs
\end